\documentclass[12pt,preprint]{aastex}
\usepackage{epsfig}
                                                                                
\def\vkm{km s$^{-1}$}

\def\degree{$^\circ$}
\def\arcs#1{$#1''$}
\def\arcsa#1#2{$#1^{\prime\prime}_{^\textrm{.}}#2$}

\def\solarmass{$M_\odot$}
\def\Jupmass{$M_\textrm{\scriptsize Jup}$}

\def\mJyb{mJy beam$^{-1}$}

\def\mJybk{mJy beam$^{-1}$ km s$^{-1}$}

\def\cmc{cm$^{-3}$}
\def\cms{cm$^{-2}$}
\def\micron{$\mu$m}
\def\VLSR{V_\textrm{\scriptsize LSR}}
\def\Vsys{V_\textrm{\scriptsize sys}}
\def\Voff{V_\textrm{\scriptsize off}}

\def\ra#1#2#3#4{#1^\mathrm{h} #2^\mathrm{m} #3^\mathrm{s}_{^\textrm{.}} #4}
\def\dec#1#2#3#4{#1\degr #2\arcmin #3^{\prime\prime}_{^\textrm{.}}#4}

\def\mH2{m_{\textrm{\scriptsize H}_2}}

\def\H2{H$_2$}
\def\N2HP{N$_2$H$^+$}
\def\HCOP{HCO$^+$}

\def\NH3{NH$_3$}

\def\SOt{$N_J=8_9-7_8$}

\def\HCOP{HCO$^+$}
\def\aHCOP{H$^{13}$CO$^+$}
                                                                                
\def\putfiga#1#2#3{\epsfig{scale=#1,angle=#2,figure=#3}}
\def\putfig#1#2#3{}
\def\leftblank#1{}

\begin{document}

\title{ALMA observations of the very young Class 0 protostellar system HH
211-mms: a 30-au dusty disk with a disk-wind traced by SO?}

\author{Chin-Fei Lee\altaffilmark{1,2}
, Zhi-Yun Li\altaffilmark{3}, Naomi Hirano\altaffilmark{1}, Hsien Shang\altaffilmark{1},
Paul T.P.  Ho\altaffilmark{1,4}, and Qizhou
Zhang\altaffilmark{5}}

\altaffiltext{1}{Academia Sinica Institute of Astronomy and Astrophysics,
P.O. Box 23-141, Taipei 106, Taiwan; cflee@asiaa.sinica.edu.tw}
\altaffiltext{2}{Graduate Institute of Astronomy and Astrophysics, National Taiwan
   University, No.  1, Sec.  4, Roosevelt Road, Taipei 10617, Taiwan}
\altaffiltext{3}{Astronomy Department, University of Virginia, Charlottesville, VA 22904, USA}
\altaffiltext{4}{East Asian Observatory, 660 N. A'ohoku Place, University Park, Hilo, HI 96720, USA}
\altaffiltext{5}{Harvard-Smithsonian Center for Astrophysics, 60 Garden
Street, Cambridge, MA 02138, USA}

\begin{abstract}

HH 211-mms is one of the youngest Class 0 protostellar systems in Perseus at
$\sim$ 235 pc away.  We have mapped its central region at up to $\sim$ 7 AU
(\arcsa{0}{03}) resolution.  A dusty disk is seen deeply embedded in a
flattened envelope, with an intensity jump in dust continuum at $\sim$ 350
GHz.  It is nearly edge-on and is almost exactly perpendicular to the jet
axis.  It has a size of $\sim$ 30 au along the major axis.  It is
geometrically thick, indicating that the (sub)millimeter light emitting
grains have yet to settle to the midplane.  Its inner part is expected to
have transformed into a Keplerian rotating disk with a radius of $\sim$ 10
au.  A rotating disk atmosphere and a compact rotating bipolar outflow are
detected in SO \SOt{}.  The outflow fans out from the inner disk surfaces
and is rotating in the same direction as the flattened envelope, and hence
could trace a disk wind carrying away angular momentum from the inner disk. 
From the rotation of the disk atmosphere, the protostellar mass is estimated
to be $\lesssim$ 50 \Jupmass{}.  Together with results from the literature,
our result favors a model where the disk radius grows linearly with the
protostellar mass, as predicted by models of pre-stellar dense core
evolution that asymptotes to an $r^{-1}$ radial profile for both the column
density and angular velocity.

\end{abstract}


\section{Introduction}

In the earliest phase of star formation, magnetic field may remove the
angular momentum in a collapsing core efficiently through magnetic braking,
potentially preventing a Keplerian rotating disk from forming around a
protostar \citep{Allen2003,Mellon2008}.  However, since Keplerian disks have
been detected as early as in the Class 0 phase, it is unclear how efficient
the magnetic braking can actually remove the angular momentum.  
In particular, a few Class 0 protostellar systems,
e.g., VLA 1623 \citep{Murillo2013}, L1527 \citep{Sakai2014}, and
HH 212 \citep{Lee2017disk,Lee2017com}, have been found to harbor Keplerian disks. 
Here by mapping the much younger system HH 211-mms 
with the Atacama Large Millimeter/submillimeter Array (ALMA),
we aim to determine how early and how small a disk
can form in order to provide stronger constraints on current models of disk
formation.  In addition, by comparing our disk result to those in the
literature, we can also study the disk growth in the earliest phase of star
formation.

Moreover, it is still unclear how angular momentum can be partly removed
from the disk in order for the disk material to transport from the outer
part to the inner part.  One possible mechanism is through a low-velocity
extended tenuous disk wind \citep{Konigl2000}.  Rotating molecular
outflows have been detected extending out from the disks in a few
protostellar systems
\citep{Launhardt2009,Agra-Amboage2011,Greenhill2013,Bjerkeli2016,Hirota2017,Tabone2017,Lee2018}.
If those outflows really trace the disk winds that facilitate the disk
evolution, we expect to detect a rotating molecular outflow in HH 211-mms as
well.

HH 211-mms is a very young and very low-mass Class 0 protostellar system in
Perseus \citep{Gueth1999,Froebrich2003,Rebull2007,Lee2009,Lee2010}, which is
located at a distance of $\sim$ 235 pc \citep{Hirota2008}.  It was not
detected at 24 \micron{} with {\it Spitzer} \citep{Rebull2007} and thus
appears to be very similar to the PACS Bright Red Sources (PBRS)
\citep{Stutz2013}, which seem to be among the youngest protostars in Orion. 
Therefore, it appears to be among the youngest Class 0 protostars in
Perseus.  A collimated jet and a collimated outflow are seen associated with
the central protostar
\citep{McCaughrean1994,Gueth1999,Hirano2006,Palau2006,Lee2007}.  Based on
current jet-launching models, a Keplerian disk is required to launch the
jet.  Recent observations with the Karl G.  Jansky Very Large Array (VLA)
have also shown a marginally resolved disk with a deconvolved size of $\sim$
22 au at 36.9 GHz at $\sim$ 16 au (\arcsa{0}{07}) resolution
\citep{Segura-Cox2016}.  In this paper, we zoom in to the innermost region
of this system with ALMA at up to 7 au (\arcsa{0}{03}) resolution at 350
GHz.  We not only confirm the presence of the disk but also spatially
resolve it in dust continuum along the major and minor axis.  We also detect
a rotating disk atmosphere and a compact rotating molecular outflow in SO. 
We will discuss the disk formation and check if the rotating outflow can
trace a disk wind.  We will also discuss how fast a disk can grow by
comparing our disk result to those in the literature.

\section{Observations}\label{sec:obs}

We have mapped HH 211-mms (Project ID: 2015.1.00024.S) with ALMA at $\sim$
350 GHz in Band 7, with one pointing toward the center.  As shown in Table
\ref{tab:obs}, C36-7 array configuration was used with 37-41 antennas.  Four
executions were carried out in 2015 in Cycle 3, one on December 4, one on
December 5, and two on December 6.  The execution on December 5 was excluded
because of its poor phase stability due to high water vapor column density. 
The projected baselines are 15-7930 m.  The resulting maximum recoverable
scale is $\sim$ \arcsa{0}{4}.  As shown in Table \ref{tab:corr3}, we had 6
spectral windows in the correlator setup, with five for molecular lines at a
velocity resolution of $\sim$ 0.212 \vkm{} per channel and one for continuum
(with possible weak lines) at a velocity resolution of $\sim$ 0.848 \vkm{}
per channel.  The total usable time on HH 211-mms is $\sim$ 123 minutes.

The CASA package was used to calibrate the data.  The bandpass, flux, and
phase calibrators are listed in Table \ref{tab:calib} with their flux
densities.  Although there was a jump in the flux density of the flux
calibrator from 2015-12-04 to 2015-12-05, the resulting flux densities of
the phase calibrator are the same and the resulting flux densities of HH
211-mms are consistent to within $\sim$ 10\%.  Here, we only present the
results in continuum and SO, with the continuum tracing the dusty disk and
SO tracing the disk atmosphere and rotating bipolar outflow.  Previously,
\HCOP{} line was thought to trace a disk \citep{Lee2009}.  However, here at
high resolution, we see that it suffers from an absorption against the disk
continuum and thus no emission is detected toward the disk, as seen in HH
212 \citep{Lee2017com}.  Therefore, \HCOP{} line can only trace the
flattened envelope around the disk and thus will be reported in the near
future.  No clear \aHCOP{} line is detected.  SiO and CO lines trace the jet
and thus will be reported in the near future.

Line-free channels were combined to produce the continuum channels.  Robust
weighting factors of 0.5 and $-$0.5 were used for the visibility to generate
the continuum maps at $\sim$ \arcsa{0}{047}$\times$\arcsa{0}{031} resolution
with a noise level of 0.11 \mJyb{} (or 0.73 K) and
\arcsa{0}{036}$\times$\arcsa{0}{024} resolution with a noise level of 0.154
\mJyb{} (or 1.75 K), respectively.  On the other hand, since SO emission is
weak, a robust weighting factor of 2 was used for the visibility to generate
the SO maps at $\sim$ \arcsa{0}{064}$\times$\arcsa{0}{046} resolution.  In
order to achieve a better detection, we binned the channel maps to have a
velocity resolution of 3 \vkm{} and a noise level of $\sim$ 1.6 \mJyb{} (or
5.6 K).  In the channel maps, the velocities are in the LSR system.

\section{Results}

In HH 211-mms, the systemic velocity is $\Vsys= 9.2\pm0.1$
\vkm{} LSR, as derived from the optically thin line of \aHCOP{} toward the
system \citep{Gueth1999}.  Throughout this paper, an offset velocity $\Voff
= \VLSR - \Vsys$ is defined to simplify our presentations.  In this
system, the jet has a position angle of 116.5\degree{} and an inclination
angle of $\lesssim$ 6\degree{} to the plane of the sky, with the
southeastern component tilted toward us \citep{Gueth1999,Lee2009,Jhan2016}.

\subsection{Disk in Continuum at 350 GHz} \label{sec:env}

\begin{figure} [!hbp]
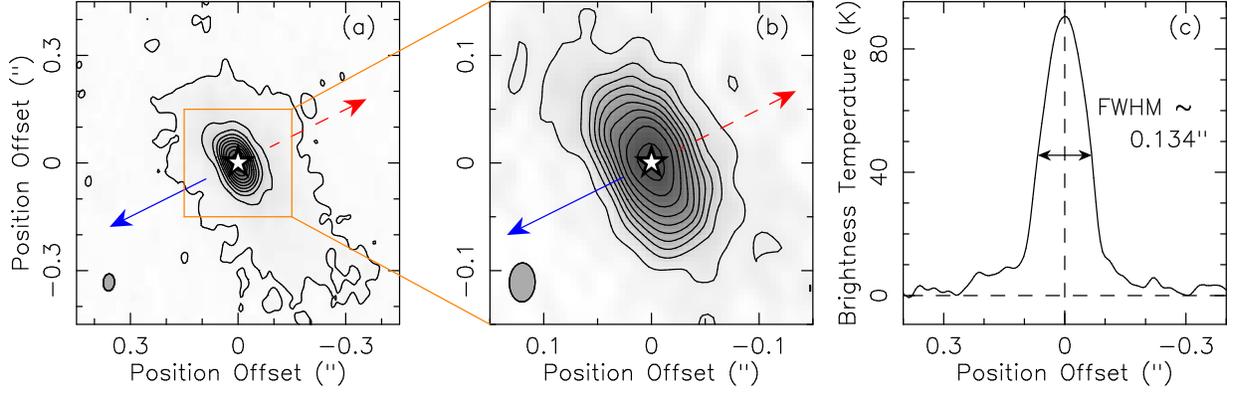

\centering
\putfiga{0.65}{270}{f1.eps} 
\figcaption[]
{ALMA continuum maps toward the center of the HH 211-mms system at 350 GHz.
The star marks the position of the central protostar.
The dashed and solid arrows indicate the axes of the redshifted component
and blueshifted component of the jet, respectively. 
Panel (a) shows the continuum map at a resolution of \arcsa{0}{047}$\times$\arcsa{0}{031}.
The contours start at 5$\sigma$ with a step of 12$\sigma$, where $\sigma=0.73$ K. 
Panel (b) shows the continuum map at a resolution of \arcsa{0}{036}$\times$\arcsa{0}{024}.
The contours start at 5$\sigma$ with a step of 5$\sigma$, where $\sigma=1.75$ K. 
Panel (c) shows an intensity cut of the continuum emission along the major axis of the envelope-disk, 
extracted from the map in panel (b).
\label{fig:cont}}
\end{figure}

Figure \ref{fig:cont}a shows the continuum map toward the central region at
the frequency of $\sim$ 350 GHz at a resolution of $\sim$ \arcsa{0}{04}.  A
compact and bright disklike structure is seen inside an extended and faint
envelope that is elongated roughly perpendicular to the jet axis.  The
envelope and disk have a total flux density of $\sim$ 192$\pm38$ mJy,
measured within the 5$\sigma$ contour. Notice that unlike the previous
continuum map shown at $\sim$ \arcsa{0}{2} resolution in \citet{Lee2009}, no
secondary emission peak is seen here at $\sim$ \arcsa{0}{3} toward the
southwest.  In order to confirm this, we made a continuum map with a similar
resolution to that in \citet{Lee2009} using a taper of \arcsa{0}{25}, but
still only saw a smooth extension of the envelope toward the southwest to
$\sim$ \arcsa{0}{4}.  Thus, the secondary emission peak seen in
\citet{Lee2009} is likely an artifact due to limited uv-coverages of the
Submillimeter Array when using the super-uniform weighting on the
visibilities. Zooming into the center at a higher resolution of $\sim$
\arcsa{0}{03}, we can better resolve the disk structure, as shown in Figure
\ref{fig:cont}b.  Notice that the envelope becomes mostly below 3$\sigma$
detection due to higher noise level in brightness temperature at higher
resolution.  The disk has an emission peak at the position
$\alpha_{(2000)}=\ra{3}{43}{56}{8054}$ and
$\delta_{(2000)}=\dec{32}{00}{50}{189}$, which is considered to be the
location of the central protostar in this paper.  The disk has a total flux
density of $\sim$ 83$\pm16$ mJy,  measured within $\sim$ 5$\sigma$
contour.

A two-dimensional Gaussian fit to the disk structure indicates that the disk
has a major axis at a position angle of 27.6\degree{} and is thus almost
exactly perpendicular to the jet axis.  This, together with the known jet
orientation, suggests that the disk is close to edge-on, with an inclination
angle of $\lesssim$ 6\degree{} to the plane of the sky (here 0\degree{} for
an edge-on disk) and the nearside tilted to the northwest.  The disk has a
deconvolved size (FWHM) of \arcsa{0}{127} (or $\sim$ 30 au) along the major
axis and \arcsa{0}{067} ($\sim$ 16 au) along the minor axis.  The aspect
ratio of the disk in the plane of the sky is $\sim$ 0.5, larger than that
expected for a thin disk with an inclination angle of 6\degree{}, which is
$\sim$ 0.1.  Therefore, the disk is geometrically thick, as found in the
slightly evolved but vertically resolved edge-on Class 0 disks, e.g.,
HH 212 \citep{Lee2017disk}.  This indicates that the (sub)millimeter light
emitting grains have yet to settle to the midplane.  This is different from
the Class I disk HH 30, which is very thin in dust continuum image
(unresolved vertically, Menard et al.  in prep), or the Class I/II disk HL
Tau, which has indication for dust settling from the shape of the gaps
\citep{Brogan2015}.  Observations at higher resolution are needed to check
in HH 211-mms for a dark lane similar to that seen in HH 212
\citep{Lee2017disk}.

An intensity cut along the major axis of the disk
shows an intensity jump at the position offsets of $\sim \pm$\arcsa{0}{1}
from $\sim$ 10 K there to $\sim$ 90 K at the source position (Fig. 
\ref{fig:cont}c).  These position offsets can be considered as the locations
where the innermost envelope transitions to the disk, as discussed in HH 212
\citep{Lee2017disk,Lee2017com}.  The intensity profile has a FWHM $\sim$
\arcsa{0}{134}, resulting in a deconvolved FWHM of \arcsa{0}{127} along the
major axis, as found earlier in the two-dimensional Gaussian fit to the disk
structure.



We can study the disk properties by estimating the spectral index $\alpha$
(with the flux density $F_\nu\propto \nu^\alpha$) in the spectral energy
distribution of the disk emission, using the flux densities of the disk
emission at two different frequencies.  Previously, the disk has been
detected by the VLA at 36.9 GHz at a resolution of $\sim$ \arcsa{0}{07} with
a flux density of $\sim$ 0.855 mJy \citep{Tobin2016,Segura-Cox2016}.  Part
of this flux density is believed to be from free-free emission, because the
spectral index between 36.9 GHz and 28.5 GHz is $\sim$ 1.65
\citep{Tobin2016}.  Recent analysis shows that the flux density of the
dust emission corrected for the free-free emission is $\sim$ 0.57 mJy at
$\sim$ 33 GHz ($\lambda \sim$ 9 mm) in Ka-band observations
\citep{Tychoniec2018}.  Therefore, with the flux density of the disk
measured above at 350 GHz, we have $\alpha \sim$ 2.11$\pm0.2$, slightly
higher than that for an optically thick thermal dust emission, for which
$\alpha=2$.  Thus, the disk here must be warm and partly optically thick.



Assuming that the disk
emission is optically thin, the (gas and dust) mass of the disk can be roughly
estimated with the following formula 
\begin{eqnarray} 
  M_\textrm{\scriptsize disk} \sim \frac{D^2 F_\nu}{B_\nu(T_{dust}) \kappa_\nu}
\end{eqnarray} 
where $D$ is the distance to the source, $B_\nu$ is the blackbody intensity
at the dust temperature $T_{dust}$, $F_\nu$ is the observed flux density,
and $\kappa_\nu$ is the mass opacity per gram of gas and dust mass.  As
discussed earlier, the observed brightness temperature reaches 90 K, thus we
assume $T_{dust}=100$ K.  Since the mass opacity is uncertain, we assume two
different mass opacities for two different phases of star formation.  One is
the empirical opacity derived from T-Tauri disks in the late phase of star
formation, which is $\kappa_\nu = 0.1 (\nu/10^{3} \textrm{GHz})^\beta$
cm$^2$ g$^{-1}$ \citep{Beckwith1990}, and the other is the mass opacity of
protostellar cores in the early phase of star formation
\citep{Ossenkopf1994},  which is $\kappa_\nu = 0.00899 (\nu/231
\textrm{GHz})^\beta$ cm$^2$ g$^{-1}$ \citep{Tychoniec2018}, where $\beta$
is the dust opacity index.  The actual opacity of the HH 211 protostellar
disk can be in between the two.  We assume $\beta=1$, as usually assumed for
protostellar disks \citep{Andrews2009}.  The resulting disk masses are $\sim
1.8$ \Jupmass{} and $\sim 4.6$ \Jupmass{}, respectively.  Since the emission
is at least partially optically thick, the masses estimated here are only
lower limits.

We can compare the disk masses estimated above with that derived from
the emission at 33 GHz, which is more optically thin.  At that frequency,
the mass opacity is an extrapolation from the above mass opacities (which
were derived for the wavelength $\lambda \lesssim 1.3$ mm) and is thus more
uncertain.  Adopting a flux density of 0.57 mJy at 33 GHz
\citep{Tychoniec2018}, the resulting disk masses are $\sim$ 14 \Jupmass{}
and 35 \Jupmass{}, respectively, for the two different mass opacities, and
thus are much higher than those estimated above.  Previously with the same
formula, \citet{Tychoniec2018} has derived a disk mass of $\sim$ 120
\Jupmass{}, assuming a dust temperature of 30 K and a mass opacity of
protostellar cores.  As discussed above, a dust temperature of $\sim$ 100 K
is more reasonable.  Thus the disk mass should be revised to be $\sim$ 35
\Jupmass{}.  Since the mass opacity is so uncertain and the continuum at
lower frequency can probe further in where the temperature is higher,
further work is needed to check the disk masses derived here at 33 GHz.










\subsection{Disk Atmosphere and Rotating Outflow in SO}\label{sec:SO}

\begin{figure} [!hbp]
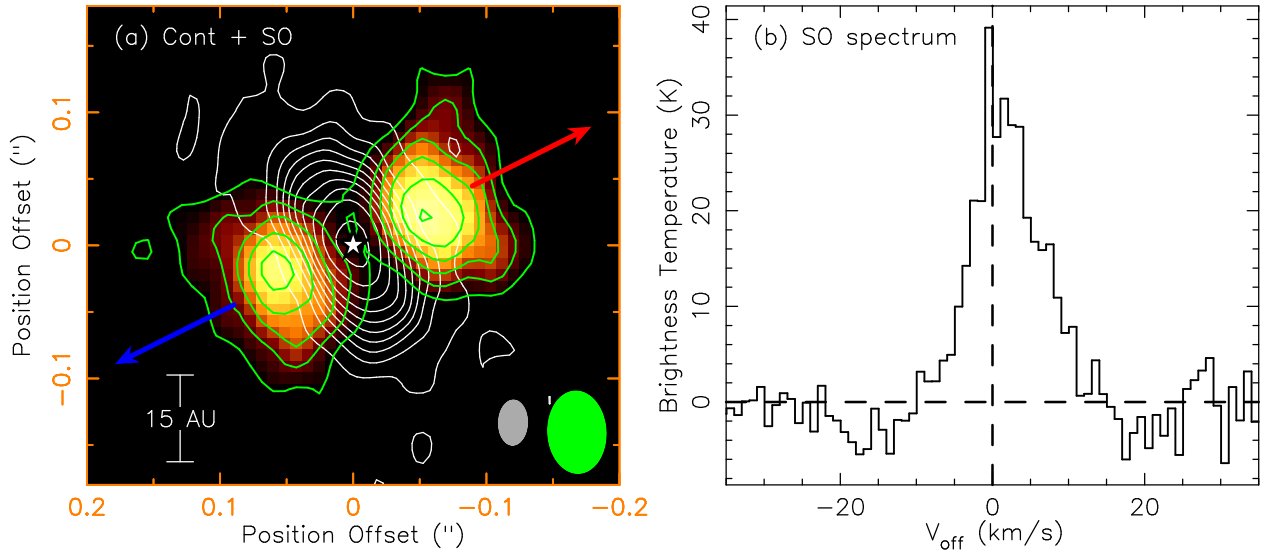

\centering
\putfiga{0.67}{270}{f2.eps} 
\figcaption[]
{SO results toward the center of the HH 211-mms system.
Panel (a) shows the SO map (in color and contours) on top of the continuum map. 
The beam has a size of \arcsa{0}{064}$\times$\arcsa{0}{046}.
The contours start at 3$\sigma$ with a step of 2$\sigma$, 
where $\sigma=15$ \mJybk{} (or $\sim$ 52 K \vkm{})
Panel (b) shows the SO spectrum averaged over a rectangular region of
\arcsa{0}{25}$\times$\arcsa{0}{12} aligned with the jet axis. 
\label{fig:SO}}
\end{figure}

SO \SOt{} emission is detected within $\sim$ \arcsa{0}{13} (30 au) of the
central source, as shown in Figure \ref{fig:SO}a.  It traces a compact
bipolar outflow, consisting of a southeastern (SE) component and a
northwestern (NW) component, extending out from the disk surfaces
surrounding the jet axis.  The outflow width reaches $\sim$ \arcsa{0}{2} (46
au) at a height of $\sim$ \arcsa{0}{1} (23 au).  Note that faint SO emission
is also detected along the jet axis at a distance $>$ \arcs{1} from the
central source, tracing the knots in the jet as seen before in
\citet{Lee2010}, but it is too faint to be discussed here meaningfully. 
Almost no SO emission is detected along the major axis of the dusty disk
likely because the dust emission there is optically thick and bright, as
seen in HH 212 \citep{Lee2017com}.  As shown in Figure \ref{fig:SO}b, the
spectrum averaged over the emitting region shows that most emission is
within $\sim$ $\pm$10 \vkm{} of the systemic velocity.  The spectrum profile
is asymmetric about the systemic velocity, with less emission in the blue.

\begin{figure} [!hbp]
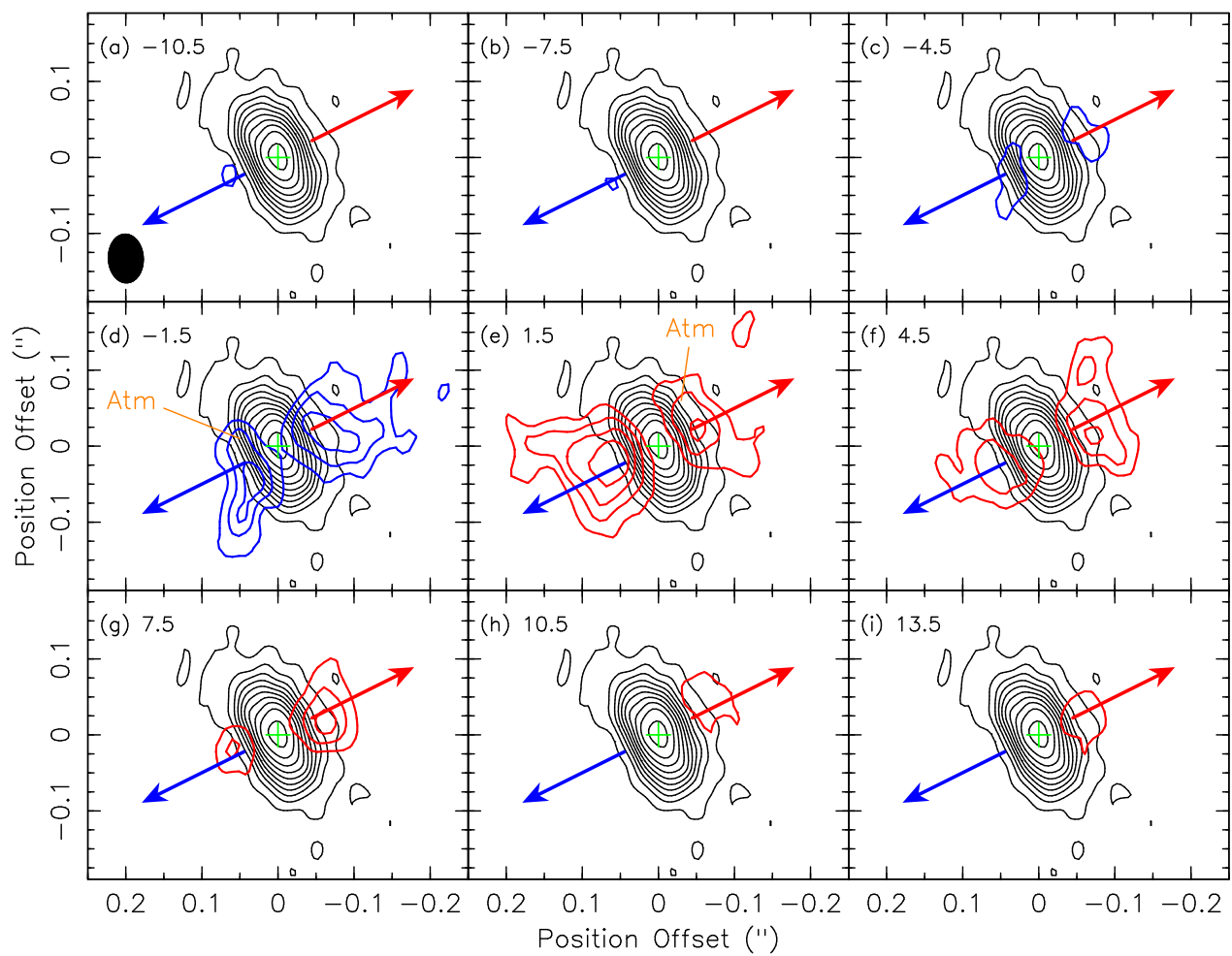

\centering
\putfiga{0.75}{270}{f3.eps} 
\figcaption[]
{Channel maps of the SO emission plotted on top of the continuum map.
The contours start at 3$\sigma$ with a step of 2$\sigma$, where $\sigma=1.6$ \mJyb{}. 
The velocity $\Voff$ in \vkm{} is labeled in the upper left corner of each panel.
In panels (d) and (e), ``Atm" means atmosphere.
\label{fig:SOchan}}
\end{figure}

The structure of the SO emission can be better seen in the channel maps
shown in Figure \ref{fig:SOchan}.  The SE and NW outflow components are seen
in both redshifted and blueshifted velocities, with the SE component
extending to higher blueshifted velocities and the NW component extending to
higher redshifted velocities.  This is expected because the outflow must have
similar inclination to the jet and is thus almost in the plane of the sky,
with the SE component tilted slightly toward us and the NW component tilted
slightly away from us.  At low velocities with $|\Voff| \lesssim 4.5$
\vkm{}, the SO emission on the dusty disk surfaces forms flattened
structures aligned with the dusty disk surfaces (see Figure
\ref{fig:SOchan}c to \ref{fig:SOchan}f), and thus actually traces the disk
atmosphere, as seen in HH 212.  The disk atmosphere can be clearly seen at
the lowest velocities as pointed in Figures \ref{fig:SOchan}d and
\ref{fig:SOchan}e, with its radius estimated to be $\sim$ \arcsa{0}{05} or
12 au, using a 4$\sigma$ detection level.
 Above the disk surfaces, the SO emission forms
shell-like structures around the jet axis opening to the southeast and
northwest from the inner disk (see Figure \ref{fig:SOchan}c to
\ref{fig:SOchan}f), tracing the outflow shells.  Going from the blueshifted
to redshifted velocity, the emission of the shell moves from the southwest
to the northeast of the jet axis.  This velocity sense is the same as that
seen before in the \HCOP{} rotating flattened envelope \citep{Lee2009},
implying that the shell is rotating with the same velocity sense as the
flattened envelope.  At higher velocity with $|\Voff| > 4.5$ \vkm{}, the
emission is seen along the jet axis, probably tracing the front and back
walls of the outflow shells projected along the jet axis.

\begin{figure} [!hbp]
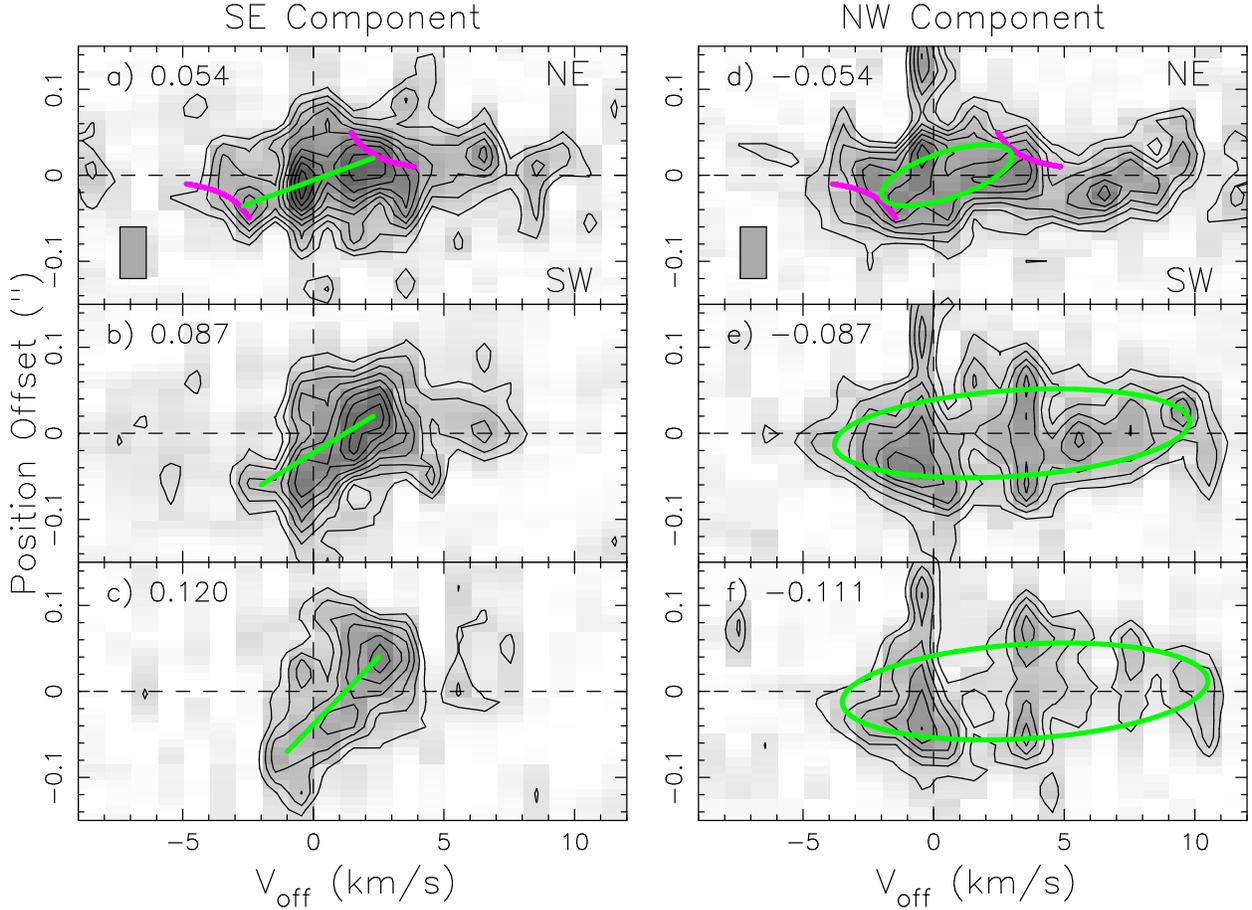

\centering
\putfiga{0.65}{270}{f4.eps} 
\figcaption[]
{PV diagrams of the SO emission cut perpendicular to the jet axis, centered
at increasing distance from the central protostar.
The contours start at 2$\sigma$ with a step of 1$\sigma$, where $\sigma=6.8$ K. 
The distance from the central protostar is given in the upper left corner in arcsecond.
The bars in the lower left corners of panels (a) and (d) 
show the velocity and angular resolutions of the PV diagrams for the SE component
and NW component, respectively.
In panels (a) and (d), the magenta curves mark the 
Keplerian rotation curves due to a central protostar with a
mass of $\sim$ 50 \Jupmass{}.
In panels (b) and (c), the green linear lines show the possible
velocity gradient of the outflow shell in the SE outflow component.
In panels (e) and (f), the tilted green ellipses show the possible
PV structure of the outflow shell in the NW outflow component.
\label{fig:pvso}}
\end{figure}

Figure \ref{fig:pvso} shows the position-velocity (PV) diagrams across the
jet axis at increasing distance from the central source.  The kinematics of
the disk atmosphere can be studied with the PV diagrams at the disk surfaces
in Figures \ref{fig:pvso}a \& \ref{fig:pvso}d.  As mentioned above, the disk
atmosphere is traced by the emission on the disk surfaces with $|\Voff|
\lesssim 4.5$ \vkm{}.  There the redshifted emission is seen in the
northeast and the blueshifted in the southwest (see Figures \ref{fig:pvso}a
\& \ref{fig:pvso}d), the same as those seen in the \HCOP{} rotating
flattened envelope.  Notice that in Figure \ref{fig:pvso}d, the emission at
$\Voff \sim -$0.5 \vkm{} and \arcsa{0}{12} is a separate component not
associated with the disk atmosphere in the northwest, and further
observations are needed to determine its origin.  In addition, the outer
boundaries of the PV structures can be roughly outlined by the magenta
curves, which are the Keplerian rotation curves due to a protostellar mass
of 50 \Jupmass{}, suggesting that the central protostar has a mass of $\lesssim$ 50
\Jupmass{}.  For the disk atmosphere in the southeast, a linear PV structure
(as delineated by the green line in Figure \ref{fig:pvso}a) is seen,
revealing a rotating ring in the atmosphere.  For the disk atmosphere in the
northwest, a roughly elliptical-like PV structure (as delineated by the
tilted green ellipse in Figure \ref{fig:pvso}d) can be seen, revealing a
rotating and expanding ring in the atmosphere.  This indicates that the
atmosphere there has been significantly affected by the outflow, as seen in
HH 212 \citep{Lee2018}.  Notice that the center of the PV structure
(bracketed by the Keplerian curves) has a velocity shift of $\sim -0.5$
\vkm{} in the SE atmosphere and $\sim 0.5$ \vkm{} in the NW atmosphere. 
This implies that the atmosphere also has a small outflow velocity of $\sim$
5 \vkm{}, assuming an inclination angle of 6\degree{}.  However, since the
velocity shift is small, further observations are needed to confirm the
outflow velocity of the atmosphere.


The kinematics of the bipolar rotating outflow can be studied with the PV
diagrams above the disk surfaces in Figures \ref{fig:pvso}b-\ref{fig:pvso}c
\& \ref{fig:pvso}e-\ref{fig:pvso}f.  For the SE outflow component, the
blueshifted and redshifted emission are seen on the opposite sides of the
jet axis (Figures \ref{fig:pvso}b and \ref{fig:pvso}c), as discussed above. 
The specific angular momentum of the outflow is estimated to be $\lesssim$
20 au \vkm{}, with the largest possible value marked by the linear line in
the diagrams.  For the NW outflow component (Figures \ref{fig:pvso}e and
\ref{fig:pvso}f), as guided by the tilted green ellipse, the emission
structure could form a tilted elliptical PV structure, as seen before in the
rotating outflow in Orion BN/KL Source I \citep{Hirota2017} and HH 212
\citep{Lee2018}.  From these elliptical PV structures, we estimated a
similar specific angular momentum of $\sim$ 20 au \vkm{}, with an expansion
velocity of $6-7$ \vkm{}.  On the other hand, we can not identify any
elliptical PV structure in Figures \ref{fig:pvso}b and \ref{fig:pvso}c, and
hence can not estimate the expansion velocity for the SE outflow component.

\begin{figure} [!hbp]
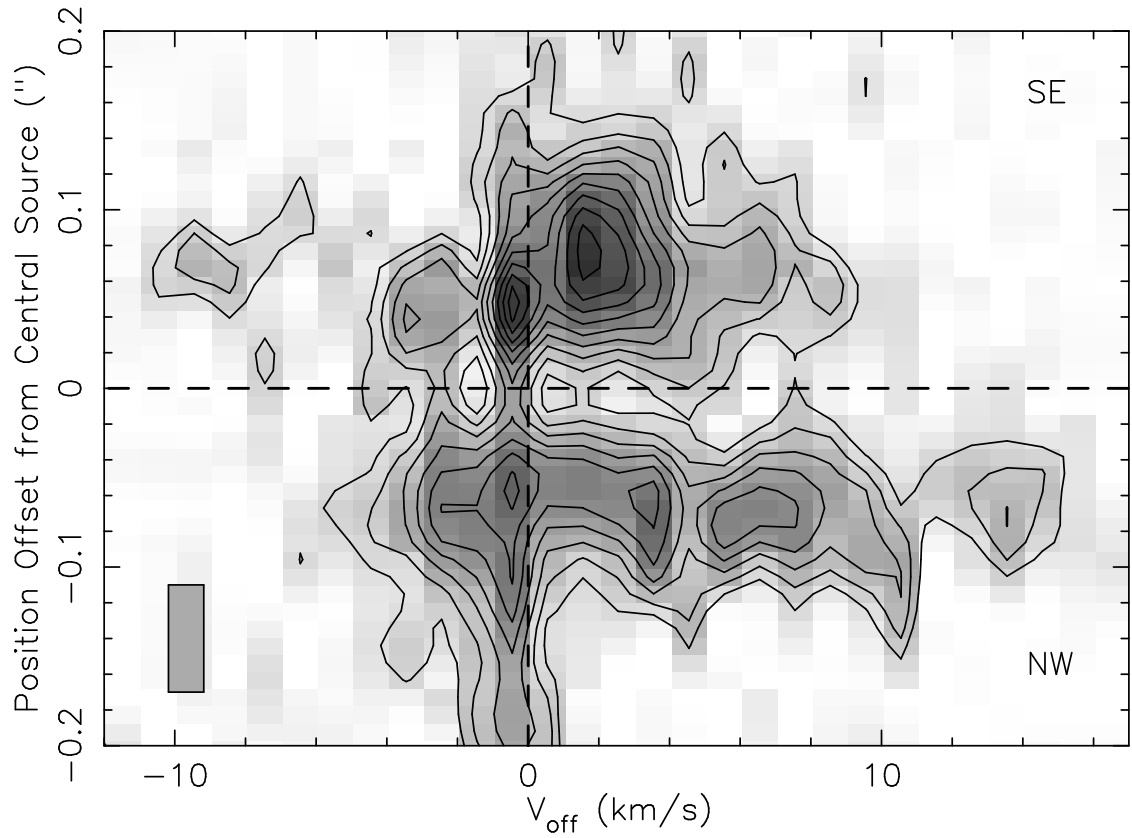

\centering
\putfiga{0.6}{270}{f5.eps}
\figcaption[]
{PV diagram of the SO emission cut along the jet axis.
The contours start at 2$\sigma$ with a step of 1$\sigma$, where $\sigma=6.8$ K. 
The bar in the lower left corner shows the velocity and angular
resolutions of the PV diagram.
\label{fig:pvsojet}}
\end{figure}

Figure \ref{fig:pvsojet} shows the PV diagram cut along the jet axis. 
Most of the emission is associated with the disk atmosphere with position
offsets of $\sim$ $\pm$\arcsa{0}{06}.  Not much emission is seen from the
outflow along the jet axis.  Future observations are needed to resolve the
PV structure of the disk atmosphere along the jet axis.


As seen in Figure \ref{fig:SO}a, the SO emission has a mean intensity of 300$\pm$60 K \vkm{}.  Assuming an
optically thin emission in local thermodynamic equilibrium and an
excitation temperature of $\sim$ 100 K as in HH 212 \citep{Lee2018}, the
mean SO column density is estimated to be $\sim$ $4.4\pm0.9\times10^{15}$
\cms{}.  Assuming an SO abundance of $2\times10^{-6}$ as found in the jet
\citep{Lee2010}, the molecular hydrogen column density is $\sim$
$2.2\pm0.4\times10^{21}$ \cms{}.   This SO abundance in the jet was
derived before from dividing SO column density by H$_2$ column density
(derived from CO emission) in the jet \citep{Lee2010}. Given that the shell
thickness is $\lesssim$ \arcsa{0}{05} or 12 au, the mean density in the
shell is $\gtrsim 1.2\times10^{7}$ \cmc{}, which is close to the critical
density of the SO line.  The total flux of the SO emission is $\sim$ 0.87 Jy
\vkm{}.  Thus, the total mass in the atmosphere and rotating outflow is
$\sim 1.3\times10^{-3}$ \Jupmass{}.


\section{Discussion}

\subsection{Disk Growth and Magnetic Braking}

In some early models of magnetized core collapse, magnetic braking can
prevent a Keplerian rotating disk from forming around a central protostar in
the earliest phase of star formation \cite[e.g.,][]{Allen2003,Mellon2008}. 
However, recent works have shown that in addition to misaligned magnetic
fields \citep{Hennebelle2009}, non-ideal MHD effects, e.g., Ohmic
dissipation and ambipolar diffusion, can enable formation of Keplerian
rotating disks \citep{Dapp2010,Machida2011,Masson2016,Zhao2018}.  Recent
ALMA observations at unprecedented angular resolution have also shown that a
Keplerian disk can form in $\sim$ $5\times10^4$ yrs old Class 0 system HH
212 \citep{Lee2017disk,Lee2017com}.  In that system, a dusty disk with a
radius of $\sim$ 60 au is detected with an intensity jump of $\sim 10$ from
the envelope to the disk in the continuum emission at 350 GHz.  This
intensity jump is believed to trace a density jump produced by an accretion
shock around the disk and the envelope transforms to a Keplerian disk after
passing through the shock.  Indeed, a Keplerian disk with a smaller radius
of $\sim$ 44 au, which is about two-third of the dusty disk radius, is
confirmed later with a kinematic study in molecular lines
\citep{Lee2017com}.


HH 211-mms is younger than HH 212.  Interestingly, a similar intensity jump
of $\sim$ 9 is also seen in the continuum emission at 350 GHz at a radius of
$\sim$ 15 au (see Section \ref{sec:SO}).  Since this intensity jump can also
trace a density jump produced by an accretion shock, a Keplerian disk might
have formed with a smaller radius of $\sim$ 10 au, roughly in agreement with
the radius of the rotating disk atmosphere detected in SO.  A similar disk
radius was also estimated at 36.9 GHz (or 8.1 mm) \citep{Segura-Cox2016}. 
In this system, the rotation axis of the disk must be aligned with the jet
axis, because the disk is almost exactly perpendicular to the jet.  Since
the magnetic field in the cloud core is mostly north-south oriented
\citep{Matthews2009,Hull2014}, a large misalignment of $\sim$ 60\degree{}
exists between the magnetic field axis in the cloud core and the rotation
axis of the disk.  Hence, it is possible that the magnetic braking here is
not efficient enough to prevent a disk from forming.  Previous dust
polarization observations toward the inner envelope of HH 211-mms showed a
possible hint of a toroidal magnetic field wrapping around the disk, further
supporting this possibility \citep{Lee2014HH211}.  The disk here is small
likely because the specific angular momentum in the envelope is small, which
is roughly estimated to be $\sim$ 35 au \vkm{} from the \HCOP{} envelope in
\citet{Lee2009}.  This value is much smaller than that in HH 212, which is
$\sim$ 140 au \vkm{}, and L1527 (54 au disk), which is $\sim$ 130 au \vkm{}
\citep{Ohashi2014}.  In addition, HH 211-mms is much younger than HH 212 and
L1527, and thus it is likely that only the inner cloud core with smaller
angular momentum has collapsed into the center.  As discussed in
\citet{Tanner2011}, the radius of the dynamical collapse in HH 211-mms could
be only $\sim$ 870 au, adjusted to the new distance of 235 pc.  As shown in
their Figure 11, this is the radius where the specific angular momentum of
the extended ammonia envelope decreases to roughly the same as that of the
inner \HCOP{} envelope.



\begin{figure} [!hbp]
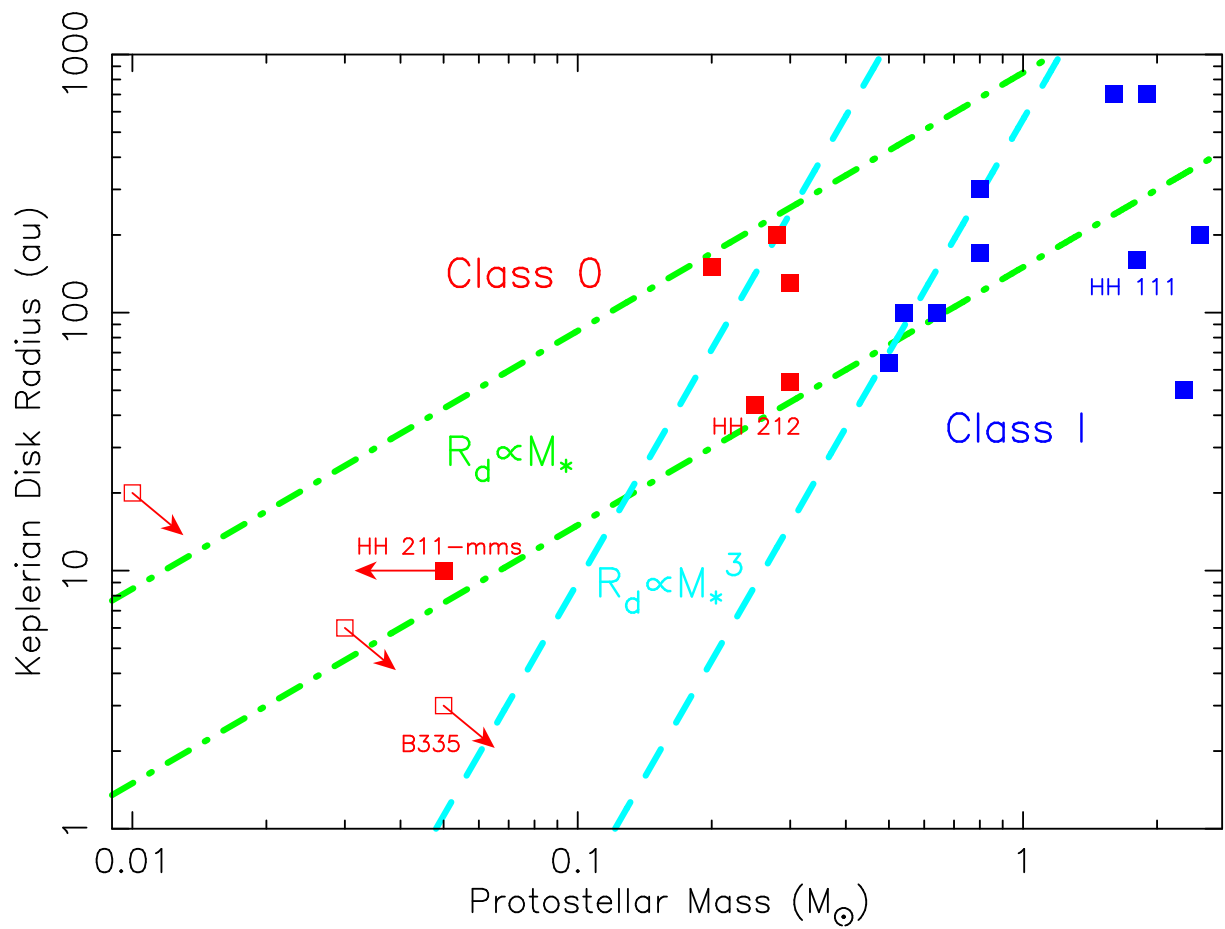

\centering
\putfiga{0.65}{270}{f6.eps} 
\figcaption[]
{A plot of protostellar mass vs. disk radius, adopted from Figure 10 in \citet{Yen2017}
after adding our measurement of HH 211-mms and the new measurement of HH 212 in \citet{Lee2017com}.
Red boxes, either filled or hollow, are for Class 0 sources. The hollow boxes
indicate the values inferred from unresolved observations, representing the upper limits.
Blue boxes are for Class I sources.  As in \citet{Yen2017}, two possible trends of disk growth are plotted, one
with green dashed lines for $R_d \propto M_\ast$, the other with cyan 
dashed lines for $R_d \propto M_\ast^3$
\label{fig:diskmass}}
\end{figure}

The growth of Keplerian disk radius from Class 0 to I phase has been studied
by \cite{Yen2017}.  Figure \ref{fig:diskmass} shows their plot after adding
our measurement of HH 211-mms and the new measurement of HH 212 in
\citet{Lee2017com}.  Previously, the trend of the disk growth with the
protostellar mass in the early Class 0 phase could not be determined
because no disk measurement was confirmed for the protostellar mass less
than 0.2 \solarmass{}.  In particular, the two small disk radii with the
protostellar mass $<0.04$ \solarmass (the two leftmost open squares) were
inferred from the observations at a resolution of $\sim$ 100 au and thus are
quite uncertain.  For B335 which has a protostellar mass of $\lesssim$ 0.05
\solarmass{}, a disk radius of $\sim$ 3 au was inferred from an observation
at $\sim$ 50 au resolution, and thus is also very uncertain.  In addition,
the very small disk radius of B335 could result from a magnetic braking
\citep{Yen2015}.  Now with our measurement of HH 211-mms, we can see more
clearly that the disk radius is large enough to be measured with
current instrument (ALMA) in the earliest phase of star formation.  The
significance of this radius measurement can be gauged from Figure
\ref{fig:diskmass}, where we present the expectations for two competing
scenarios of protostellar disk growth.  The first is based on the classic
picture of \citet{Terebey1984}, who considered the growth of a disk in the
collapse of a singular isothermal sphere (SIS) with a solid-body rotation. 
In this case, the disk radius stays small initially, but grows rapidly with
stellar mass at later times, as $R_d \propto M_*^3$.  The reason for such a
steep dependence is that the materials at small radii of the SIS rotates
very slowly and, when they collapse, they form a very small disk initially. 
The disk size grows rapidly at later times, as the materials at larger radii
of the SIS with much larger specific angular momenta collapse to the central
region.  The coefficient in front of the $M_*^3$ scaling depends on the
isothermal sound speed and especially the initial rotation rate of the SIS,
which is uncertain.  In Figure \ref{fig:diskmass}, there are two dashed
(cyan) lines for two coefficients chosen to roughly bracket the currently
available data on late Class 0 and early Class I sources.  Extension of
these lines to earlier times when the protostellar mass is smaller
represents the expectation of this ``slow-start, rapid-growth" scenario for
the youngest Class 0 disks based on the current data.  The upper limit
inferred for the disk radius for B335 and the lower limit inferred for its
stellar mass appears to provide some support for this scenario.  However,
the support is rather weak, because the disk in B335 has never been
detected.  Without detecting the disk, it would be difficult to infer the
stellar mass or put a strong upper limit on it.

An alternative scaling was advocated by \citet{Basu1998}, based on the
asymptotic state of the pre-stellar evolution of a rotating, magnetized
dense core, which is expected to have a power-law distribution for both the
column density $\Sigma \propto r^{-1}$ and the angular speed $\Omega \propto
r^{-1}$ right before the formation of a central stellar object.  Compared to
the SIS rotating as a solid body in the first scenario, the materials at
small radii rotate much faster; they collapse to form a much larger disk
compared to the first scenario.  Such a disk grows more slowly with the
stellar mass at later times, as $R_d\propto M_*$, because the specific
angular momentum in the pre-collapse configuration increases with radius
more slowly compared to the first scenario.  The coefficient in front of the
linear scaling depends on the sound speed, magnetic field strength, and the
rotation rate, and can be different for different sources.  In Figure
\ref{fig:diskmass}, two (green) dot-dashed lines with different coefficients
are plotted to roughly bracket the current data for the late Class 0 and
early Class I sources.  Extension of these lines to earlier times represents
the expectation of this ``early-start, slow-growth" scenario for the youngest
Class 0 disks.  Our measurement of the radius of HH 211-mms disk and
estimate for its mass lie in the expected region, which provides support for
this scenario; they are not consistent with the expectation of the
``slow-start, rapid-growth" scenario based on \citet{Terebey1984}. 
Indeed, the trend of linear growth of disk
radius with protostellar mass appears to extend to the late Class I
phase.  However, the disk growth in the late Class I phase could be modified
by magnetic braking, as suggested by our previous observation of HH 111
\citep{Lee2016}.



\subsection{Disk Atmosphere and Disk Wind?}

In SO, a disk atmosphere is seen on the dusty disk surfaces and a bipolar
rotating molecular outflow is seen coming out from the inner disk, as in HH
212 \citep{Lee2017com}.  The disk atmosphere has a radius of $\sim$ 12 au,
similar to the expected radius of a Keplerian disk around the central
protostar.  The SO emission also shows a warm ring in the disk atmosphere. 
Such a warm SO ring has also been detected before in other protostellar
systems, e.g., HH 111 \citep{Lee2016} and L1527 \citep{Sakai2017}, probably
tracing an accretion shock created by a rapid decrease of the infall
velocity near the centrifugal barrier where a Keplerian disk is formed.  The
warm ring in the NW is also expanding, and thus can also be produced or
affected by an interaction with the rotating molecular outflow.


The rotating molecular outflow extends out to $\sim$ 30 au ($\sim$
\arcsa{0}{13}) from the inner disk, with a width increasing with the
distance.  The width reaches $\sim$ 46 au (\arcsa{0}{2}) at a height of
$\sim$ 23 au (\arcsa{0}{1}).  The outflow shell can trace back down to the
innermost part of the disk to within a few au of the protostar.  The outflow
has an expansion velocity of $\sim$ 6 \vkm{} and a specific angular momentum
of $\lesssim$ 20 au \vkm{}, although further observations with a better
sensitivity are needed to confirm these values.  Like that in HH 212 and
others, more work is needed to confirm if the bipolar outflow really traces
the wind from the disk or merely disk material pushed away by a wide-angle
wind further in \citep{Lee2010}.  In addition, for a disk wind, we expect to
see nested shells coming out from different radii of the disk.  Thus,
further observations are also needed to check if outflow shells coming out
from different radii can be detected in different density tracers.

\section{Conclusions}

We have confirmed the presence of a disk in HH 211-mms with ALMA at up to
$\sim$ 7 au resolution.  The disk is seen inside a flattened envelope with
an intensity jump in dust continuum at $\sim$ 350 GHz.  It is nearly edge-on
and is almost exactly perpendicular to the jet axis.  It has a size of
$\sim$ 30 au along the major axis and $\sim$ 16 au along the minor axis.  It
is geometrically thick, indicating that the (sub)millimeter light emitting
grains have yet to settle to the midplane.  Its inner part is expected to
have transformed into a Keplerian rotating disk with a radius of $\sim$ 10
au.  Thus, a Keplerian disk can form around protostars with a radius as
small as $\sim$ 10 au.  Our result for HH 211-mms favors a model where the
disk size grows linearly with protostellar mass $M_*$ rather than with
$M_*^3$.  In addition, we have also detected a disk atmosphere and a compact
rotating bipolar outflow in SO \SOt{}.  The disk atmosphere is rotating and
has a small outflow velocity of a few \vkm{} in the outflow direction.  It
is also expanding and could be affected by the outflow.  The outflow extends
out to $\sim$ 30 au from the disk surfaces to the southeast and northwest
around the jet axis.  It is rotating in the same direction as the flattened
envelope, and hence could trace a wind carrying away angular momentum from
the inner disk.

\acknowledgements

We thank the referee for the constructive comments on our paper.  This paper
makes use of the following ALMA data: ADS/JAO.ALMA\#2015.1.00024.S.  ALMA is
a partnership of ESO (representing its member states), NSF (USA) and NINS
(Japan), together with NRC (Canada), NSC and ASIAA (Taiwan), and KASI
(Republic of Korea), in cooperation with the Republic of Chile.  The Joint
ALMA Observatory is operated by ESO, AUI/NRAO and NAOJ.  C.-F.L. 
acknowledges grants from the Ministry of Science and Technology of Taiwan
(MoST 104-2119-M-001-015-MY3) and the Academia Sinica (Career Development
Award).  ZYL is supported in part by NSF grant AST-1313083 and AST-1716259
and NASA grant NNX14AB38G.




\def\nat{Natur}

\clearpage

\begin{table}
\small
\centering
\caption{Observation Logs}
\label{tab:obs}
\begin{tabular}{llllll}
\hline
Cycle & Date & Array &Number of &Time on target & Projected Baselines  \\
      & (YYYY-MM-DD) & Configuration & Antennas&(minutes) & (meter) \\
\hline\hline
3 & 2015-12-04 & C36-7   & 37 & 41  & 16$-$7930   \\ 
3 & 2015-12-05 & C36-7   & 41 & 41  & 13$-$5800   \\ 
3 & 2015-12-06 & C36-7   & 37 & 41  & 15$-$6200 \\ 
3 & 2015-12-06 & C36-7   & 37 & 41  & 35$-$5890  \\ 
\hline
\end{tabular}
\end{table}

\begin{table}
\small
\centering
\caption{Correlator Setup for Cycle 3 Project}
\label{tab:corr3}
\begin{tabular}{llllll}
\hline
Spectral  & Line or   & Number of & Central Frequency & Bandwidth & Channel Width\\
Window & Continuum & Channels  & (GHz)             & (MHz)     & (kHz) \\
\hline\hline
0 & SO \SOt      & 960   & 346.528 & 234.375  & 244.140  \\
1 & CO $J=3-2$      & 960   & 345.796 & 234.375  & 244.140  \\
2 & H$^{13}$CO$^+$ $J=4-3$      & 960   & 346.998 & 234.375  & 244.140  \\
3 & SiO $J=8-7$     & 960   & 347.330 & 234.375  & 244.140  \\
4 & HCO$^+$ $J=4-3$ & 1920  & 356.735 & 468.750  & 244.140  \\
5 & Continuum     & 1920  & 357.994 &1875.000  & 976.562  \\
\hline
\end{tabular}
\end{table}

\begin{table}
\small
\centering
\caption{Calibrators and Their Flux Densities}
\label{tab:calib}
\begin{tabular}{llll}
\hline
Date & Bandpass Calibrator &Flux  Calibrator & Phase Calibrator \\
(YYYY-MM-DD) & (Quasar, Flux Density) & (Quasar, Flux Density) & (Quasar, Flux Density) \\
\hline\hline
2015-12-04 & J0237+2848, 1.22 Jy & J0238+1636, 1.11 Jy & J0336+3218, 0.51 Jy  \\ 
2015-12-05 & J0237+2848, 1.19 Jy & J0238+1636, 0.85 Jy & J0336+3218, 0.51 Jy \\ 
2015-12-06 & J0237+2848, 1.26 Jy & J0238+1636, 0.85 Jy & J0336+3218, 0.51 Jy \\  
2015-12-06 & J0237+2848, 1.23 Jy & J0238+1636, 0.85 Jy & J0336+3218, 0.51 Jy \\ 
\hline
\end{tabular}
\end{table}

\end{document}